\documentclass[sigplan,nonacm]{acmart}

\usepackage[]{hyperref}
\usepackage{chngcntr}
\counterwithout{figure}{section}
\usepackage{caption}
\usepackage{graphicx, subfig, subcaption}
\usepackage{array}
\usepackage{multicol, multirow}
\usepackage{algorithm}
\usepackage{algpseudocode}
\usepackage{float}
\usepackage{pifont}
\usepackage{tikz}
\usepackage{bbm}
\newcommand{\ourdesign}{MIGRator}
\usepackage{bm}
\usepackage{xcolor}
\usepackage{geometry}
\usepackage{tabularx}
\definecolor{asparagus}{rgb}{0.55, 0.71, 0.0}

\newcommand{\squishlist}{
    \begin{list}{$\bullet$}
        { \setlength{\itemsep}{0pt}      \setlength{\parsep}{0pt}
            \setlength{\topsep}{0.5pt}       \setlength{\partopsep}{0pt}
            \setlength{\listparindent}{-2pt}
            \setlength{\itemindent}{-5pt}
            \setlength{\leftmargin}{0.5em} \setlength{\labelwidth}{0em}
            \setlength{\labelsep}{0.2em} } }
    
\newcommand{\squishend}{
\end{list}  }

\usepackage{xcolor}

\sloppypar

\usepackage{amsmath, amsfonts}

\usepackage{amssymb}

\begin{document}

\title{Improving GPU Multi-Tenancy Through Dynamic Multi-Instance GPU Reconfiguration}


\author{Tianyu Wang}
\affiliation{%
  \institution{University of Pittsburgh}
  \city{Pittsburgh}
  \state{PA}
  \country{USA}
}
\email{tiw81@pitt.edu}

\author{Sheng Li}
\affiliation{%
  \institution{University of Pittsburgh}
  \city{Pittsburgh}
  \state{PA}
  \country{USA}
}
\email{shl188@pitt.edu}

\author{Bingyao Li}
\affiliation{%
  \institution{University of Pittsburgh}
  \city{Pittsburgh}
  \state{PA}
  \country{USA}
}
\email{bil35@pitt.edu}

\author{Yue Dai}
\affiliation{%
  \institution{University of Pittsburgh}
  \city{Pittsburgh}
  \state{PA}
  \country{USA}
}
\email{yud42@pitt.edu}

\author{Ao Li}
\affiliation{%
  \institution{University of Arizona}
  \city{Tucson}
  \state{AZ}
  \country{USA}
}
\email{aoli1@arizona.edu}

\author{Geng Yuan}
\affiliation{%
  \institution{University of Georgia}
  \city{Athens}
  \state{GA}
  \country{USA}
}
\email{geng.yuan@uga.edu}

\author{Yufei Ding}
\affiliation{%
  \institution{University of California, San Diego}
  \city{San Diego}
  \state{CA}
  \country{USA}
}
\email{yufeiding@ucsd.edu}

\author{Youtao Zhang}
\affiliation{%
  \institution{University of Pittsburgh}
  \city{Pittsburgh}
  \state{PA}
  \country{USA}
}
\email{zhangyt@cs.pitt.edu}

\author{Xulong Tang}
\affiliation{%
  \institution{University of Pittsburgh}
  \city{Pittsburgh}
  \state{PA}
  \country{USA}
}
\email{tax6@pitt.edu}

\begin{abstract}
Continuous learning (CL) has emerged as one of the most popular deep learning paradigms deployed in modern cloud GPUs. Specifically, CL has the capability to continuously update the model parameters (through model retraining) and use the updated model (if available) to serve overtime arriving inference requests. It is generally beneficial to co-locate the retraining and inference together to enable timely model updates and avoid model transfer overheads. This brings the need for GPU sharing among retraining and inferences. Meanwhile, multiple CL workloads can share the modern GPUs in the cloud, leading to multi-tenancy execution. In this paper, we observe that prior GPU-sharing techniques are not optimized for multi-tenancy CL workloads. Specifically, they do not coherently consider the accuracy of the retraining model and the inference service level objective (SLO) attainment. Moreover, they cannot accommodate the overtime dynamics (e.g., inference arrival intensity) in CL execution. In this paper, we propose \ourdesign, a novel GPU reconfiguration runtime that dynamically performs GPU reconfiguration for multi-tenancy CL workloads. \ourdesign~is based on the recent NVIDIA multi-instance GPU (MIG) to mitigate resource contention and formulates the reconfiguration optimization into Integer Linear Programming (ILP) to dynamically identify, reconfigure, and allocate the GPU instances. \ourdesign~leverages the ``Goodput'' metric in the ILP objective function to consider both inference SLO attainment and model accuracy in the reconfiguration exploration. We evaluate \ourdesign~using representative multi-tenancy CL workloads. The results show our approach outperforms the state-of-the-art GPU sharing techniques (i.e., Ekya, Astraea, and PARIS) by 17\%, 21\%, and 20\%, respectively.

\end{abstract}

\maketitle 
\pagestyle{plain} 

\section{Introduction}
\label{sec:intro}

Continuous learning (CL) is one of the most popular deep learning paradigms and has received momentum in recent years~\cite{est-acc-ekya,ctn-recl,ctn-aljundi2019task,ctn-li2017learning,ctn-lismartfrz,ctn-van2019three,ctn-xu2018reinforced}. In particular, CL has the capability to continuously update the model parameters to adapt/update the model to changing environments (also called ``data drift'' in the machine learning community)~\cite{ctn-recl,recommendation_1,recommendation_2,shubha2023adainf}. Meanwhile, CL serves overtime inference requests using the updated model to yield accuracy for overtime inference requests. Thus, CL is particularly useful for applications domains such as personalized medicine~\cite{medicine_1,medicine_2,medicine_3} and recommendation systems~\cite{recommendation_1,recommendation_2}, where overtime model customization is needed for overtime inference accuracy.

There are two important aspects that can affect continuous learning performance. On the one hand, the model must be updated in a timely manner to quickly capture the data drift changes. On the other hand, there are generally strict deadlines for inference requests to meet the Service-Level Objectives (SLOs) (e.g., autonomous robot~\cite{robot_1,robot_2,robot_3,robot_4}). Therefore, a CL inference request is only considered ``valid'' if it satisfies both conditions: i) the inference output is correct and ii) the SLO requirement is met. Due to this tight interaction between model retraining and inference in CL, it is generally beneficial to co-locate the retraining and the inferences together and share the GPU in the cloud. On the one hand, it avoids expensive transfer of updated model after retraining, which can be  more than 600 $\times$, if on separate GPUs, compared to the inference time~\cite{move_1,move_2,move_5}. On the other hand, it improves the cloud GPU utilization by co-running the retraining and inference and sharing the GPU. Meanwhile, multi-tenancy is prevalent in cloud, where multiple CL workloads share the GPU. 

\begin{table}[t]
\centering
\caption{Comparison to prior works.}
  \setlength\tabcolsep{1pt} 
\scriptsize
\begin{tabular}{|c|c|c|c|c|c|}
\hline
 & \textbf{Sharing} & \textbf{Minimize} & \textbf{Inference} & \textbf{Retraining} & \textbf{Fine-grain} \\
 & \textbf{strategy} & \textbf{interference} & \textbf{dynamics} & \textbf{benefits} & \textbf{reconfiguration} \\
 
\hline
Gpulet~\cite{spatialmps-servingheterogeneousATC} & MPS & \textcolor{orange}{Partial} & \textcolor{asparagus}{Yes} & \textcolor{red}{No} &  \textcolor{red}{No} \\
\hline
INFless~\cite{spatialmps-infless} & MPS & \textcolor{orange}{Partial}  & \textcolor{asparagus}{Yes} & \textcolor{red}{No} & \textcolor{asparagus}{Yes} \\
\hline
Astraea~\cite{spatialmps-astraea} & MPS & \textcolor{orange}{Partial}  & \textcolor{asparagus}{Yes} & \textcolor{red}{No} & \textcolor{asparagus}{Yes} \\
\hline
Ekya~\cite{est-acc-ekya} & MPS & \textcolor{orange}{Partial}  & \textcolor{red}{No} & \textcolor{asparagus}{Yes} & \textcolor{red}{No} \\
\hline
PARIS~\cite{spatialmig-paris} & MIG & \textcolor{asparagus}{Yes}  & \textcolor{asparagus}{Yes} & \textcolor{red}{No} & \textcolor{red}{No} \\
\hline
\ourdesign~(ours) & MIG & \textcolor{asparagus}{Yes} & \textcolor{asparagus}{Yes} & \textcolor{asparagus}{Yes} & \textcolor{asparagus}{Yes} \\
\hline
\end{tabular}
\label{tab:comparison}
\vspace{-15pt}
\end{table}


There are three basic GPU-sharing strategies supported by modern GPUs: i) Concurrent Multiple Kernels (CMK)~\cite{spatialcmk-horus,spatialcmk-baymax,spatialcmk-prophet,spatialcmk-yang}, ii) Multi-Process Service (MPS)~\cite{nvidia-mps-docu}, and iii) Multi-Instance GPU (MIG)~\cite{nvidia-mig-docu}. The difference is that MPS partitions the GPU computing units (i.e., SMs), MIG partitions both the memory and the computing units, and CMK does not partition resources and allows contention across co-running applications. As such, CMK can potentially achieve higher resource utilization but suffer from interference. In contrast, MIG eliminates the interference but may have an imbalance in resource allocation, leading to underutilization.  We summarize prior works in Table~\ref{tab:comparison} and label which basic sharing strategy they built upon. We observe that none of the prior works are optimized for multi-tenancy CL workloads on modern GPUs. First, prior MPS-based GPU resource allocation works (e.g., Astreaea~\cite{spatialmps-astraea}, INFless~\cite{spatialmps-infless}, and Gpulet~\cite{spatialmps-servingheterogeneousATC}) leave memory shared among tenants, causing interference, especially in large CL models. Second, while some prior works allocate the resource considering the inference dynamics, i.e., inference request arrival pattern, they are not aware of retraining benefits (i.e., the accuracy improvement brought by each retraining process), which can lead to a significant amount of inference requests using the stale model. Third, Ekya~\cite{est-acc-ekya} is the state-of-the-art resource allocation approach for CL workloads. While it considers the retraining benefits, it does not accommodate the inference dynamics. Finally, most prior works only support resource reconfiguration at certain execution time stamps. This is due to the fact that they employ an exhaustive search for beneficial resource allocation (e.g., in Ekya~\cite{est-acc-ekya} and MISO~\cite{spatialmig-miso}). The search overheads prevent them from conducting reconfiguration at finer time granularity (e.g., per second basis) which is important in continuous learning~\cite{trace-alibaba,trace-microsoft}. We quantitatively compared our approach to Ekya~\cite{est-acc-ekya}, Astraea~\cite{spatialmps-astraea}, and PARIS~\cite{spatialmig-paris} in Section~\ref{experiment}.

In this paper, we propose \ourdesign, a dynamic GPU reconfiguration runtime for multi-tenancy continuous learning workloads on modern GPUs. \ourdesign~is built upon modern MIG such that the interference among co-running tasks is eliminated. Also, \ourdesign~is aware of both inference dynamics and the retraining benefits in CL workloads. This is achieved by formulating the reconfiguration into an Integer Linear Programming (ILP) problem, which leverages a metric called ${\it Goodput}$ (detailed definition in Section~\ref{sec:design_ILP_objective}) in its objective function to take into account the SLO attainment and the retraining benefits. The ILP can be solved efficiently with much lower overheads compared to an exhaustive configuration search.
Moreover, the designed ILP in \ourdesign~explores MIG reconfiguration on a finer granularity (i.e., per second basis) to determine beneficial GPC allocations. 
The main contributions are as follows. 

\squishlist{}
    \item We reveal the unique challenges of GPU sharing for multi-tenancy continuous learning workloads. Specifically, both the retaining benefits and the SLO attainments have to be taken into account when conducting the GPU reconfiguration. As we quantitatively characterized, a naive and static configuration may compromise one or both, and is not able to accommodate the inference dynamics in CL workloads.  
    \item We design \ourdesign~which formulates the reconfiguration optimization into an Integer Linear Programming (ILP) problem, such that both SLO attainment and the retaining benefits are coherently considered during reconfiguration. \ourdesign~also leverages MIG to eliminate interference and allows fine-granular reconfiguration. 
    \item We evaluate \ourdesign~using representative CL workloads and compared it against the state-of-the-art GPU sharing strategies. Specifically, \ourdesign~achieves 17\%, 21\%, and 20\% $Goodput$ compare to Ekya~\cite{est-acc-ekya}, Astraea~\cite{spatialmps-astraea}, and PARIS~\cite{spatialmig-paris}. 
\squishend{}

\section{Background and Related Works}
\label{sec:bg}

\subsection{Continuous Learning and Data Drift}
\label{sec:bg_CL}

Continuous Learning (CL) is a popular deep learning paradigm that handles overtime model refinement through retraining and overtime inference requests. It is particularly useful in application domains where exist model customization such as in personalized medicine~\cite{medicine_1,medicine_2,medicine_3} and recommendation systems~\cite{recommendation_1,recommendation_2}.
A common characteristic of these applications is that the models need to be continuously updated to accommodate and adapt to possible and potential external environment changes, which are referred to as “data drifts” in the machine learning community~\cite{est-acc-ekya, ctn-recl, ctn-li2017learning, ctn-lismartfrz}. As such, CL effectively handles data drifts and is able to maintain high accuracy for incoming inference requests over time.

There are two important events in continuous learning: i) retraining (i.e., model refinement) and ii) inferences.
Specifically, retraining is intended to update the model to handle data drifts. It utilizes retraining data that arrive periodically over time. Usually, one round of retraining is called a ``retraining window'' in continuous learning~\cite{est-acc-ekya,ctn-recl}, which is a time period within which the retraining process can be completed. 
The typical retraining window size is 200 seconds~\cite{est-acc-ekya,ctn-recl}. The retraining data are assumed to be available at the beginning of each retraining window~\cite{est-acc-ekya, ctn-recl}. Note that the updated models are only available for inference requests after the retraining procedure is completed~\cite{est-acc-ekya,ctn-recl}.
On the other hand, 
unlike the retraining process, which occurs once per retraining window, inference requests arrive continuously during the entire retraining window.

Both retraining and inference requests are crucial to satisfy continuous learning applications to achieve a high inference accuracy and high SLO (Service Level Objective) attainment.
On the one hand, retraining is crucial for updating the model so that incoming inference requests can use the updated model, resulting in a higher inference accuracy.
On the other hand, inference requests have specific SLO requirements to meet. High SLO violations are problematic in many applications~\cite{autodrive_1,autodrive_2,recommendation_1,autodrive_3,autodrive_4}.
Note that an inference request in CL is only considered valid if it both i) meets the SLO requirements and ii) return correct results.

Therefore, when inference and retraining tasks share the GPU, it is non-trivial to determine how to allocate resources (e.g., SMs and memory)  among the tasks. On the one hand, prioritizing retraining leads to better accuracy as the model is updated faster. However, this can result in more SLO violations for inference requests due to insufficient resources.
On the other hand, prioritizing inference brings lower SLO violations. However, the retraining take longer execution due to the lack of resources so that fewer inference requests can use the updated model, leading to lower accuracy as the model cannot adapt quickly to the new environment. 
Additionally, in multi-tenancy where multiple continuous learning models share the GPU, the resource requirements for retraining tasks (to ensure prompt model adaptation) and inference tasks (to guarantee SLO that can even vary over time) differ among models (see Section \ref{sec:motivation} for details). These variations make the resource allocation more challenging.

\subsection{Multi-tenant Support on GPUs}
\label{sec:bg_spatial}

\begin{figure}
    \centering  
    \includegraphics[width=1\columnwidth]{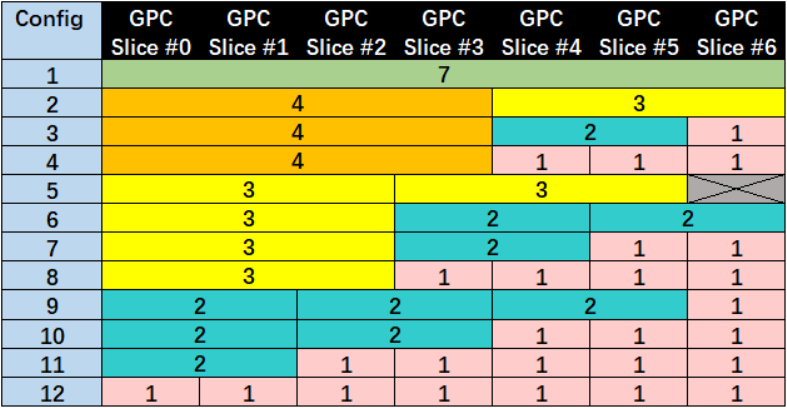}
    \caption{MIG instance configurations on NVIDIA A100.}
    \label{fig:mig_combinations}
\end{figure}

There are three popular multi-tenancy schemes on modern GPUs: i) Concurrent Mulit-Kernels (CMK), ii) Multi-Process Service (MPS), and iii) Multi-Instance GPUs (MIG). Specifically, CMK does not partition the hardware resources across tenants and relies on GPU runtime and hardware to handle the competition. As such, the interference among tenants is more significant during execution~\cite{spatialmig-miso,contention-ausavarungnirun2018mask,contention-hu2016run,contention-jog2013orchestrated}. In contrast, MPS hard partitions the computing resources (i.e., Streaming Multi-processors) among tenants while leaving the memory system shared. Finally, the MIG hard partitions both computing and memory resources across tenants to provide the strongest execution isolation. 
In modern GPU, MIG supports a total of seven Graphic Processing Clusters (GPCs), which can be dynamically combined to form GPU instances. Figure \ref{fig:mig_combinations} plots the 12 configurations supported by MIG, where each configuration consists of several instances.

\section{Motivation}
\label{sec:motivation}

We conduct three experiments to investigate the effectiveness of the MIG resource partitioning for continuous learning applications. Due to the lack of space, we only report the Inception~\cite{model-inception} and ResNet50~\cite{model-resnet50} models in this section. In fact, six different models were studied in the evaluation section, and the observations made in this section are similar in other models. The detailed experimental methodology can be found in Section~\ref{experiment_setup}.

\textbf{Dynamic resource allocation needed between retraining and inference.}
We conduct an experiment to characterize different GPC allocations between retraining and inferences for model ResNet50. We use the NC-CIFAR10 benchmark~\cite{avalanche} (details in Section~\ref{retraining_benchmarks}), which is a widely used continuous learning retraining benchmark.
In NC-CIFAR10, there are a total of 5 scenarios of training data where each scenario introduces two new data classes. The first scenario training data is used to pre-train the model, and the rest four scenario training data is used to retrain the model (corresponding to four retraining windows). 
Note that, the updated model is only available for inferences after the retraining finishes. Inference requests are received over time and served using the most updated model.
In this experiment, we measure SLO attainment and inference accuracy. SLO attainment is defined as the proportion of inference requests returned within SLO targets compared to the total number of inference requests received by the models, and inference accuracy is the proportion of inference requests that return correct results to the total number of inference requests received.

\begin{figure}[]
    \centering  
    \includegraphics[width=1\columnwidth]{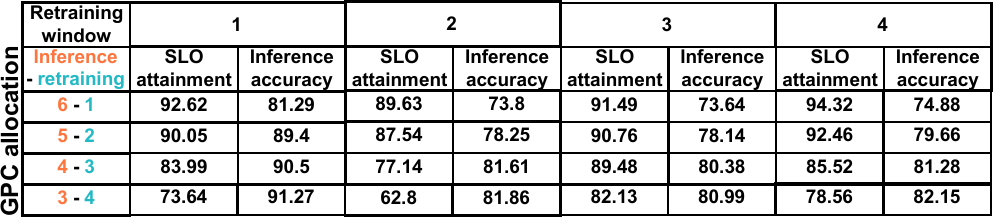}
    \caption{SLO attainment(\%) and inference accuracy(\%) of ResNet50 under various GPC allocations in four retraining windows. The leftmost column lists GPC allocations, where `6-1' indicates 6 GPCs are allocated to the inference task, and 1 GPC is allocated to the retraining task.}
    \label{fig:motivation_single_model}
\end{figure}

From Figure~\ref{fig:motivation_single_model}, first, we observe that prioritizing either inference or retraining tasks may compromise the benefits of the other one.
For example, compared to the allocation `3-4', the GPC allocation `6-1', which allocates more resources to inference tasks, achieves 18\% higher SLO attainment on average across four retraining windows but delivers an 8\% lower inference accuracy.
Second, we observe that the trade-off between the benefits of allocating more resources to retraining or inference tasks varies across retraining windows. For example, comparing the `6-1' allocation to the `3-4' allocation in window 1, we see that the `6-1' allocation yields 19\% higher SLO attainment and 10\% lower inference accuracy. In contrast, in window 2, the `6-1' allocation offers 27\% higher SLO attainment and 8\% lower inference accuracy compared to the `3-4' allocation.
The reason behind the change in trade-off is that both the inference request arrival rate and the accuracy improvement through retraining tasks vary across windows.
These two observations motivate us to coherently consider both SLO attainment and the retraining benefits when conducting resource allocation.

\begin{figure}[]
    \centering  
    \includegraphics[width=1\columnwidth]{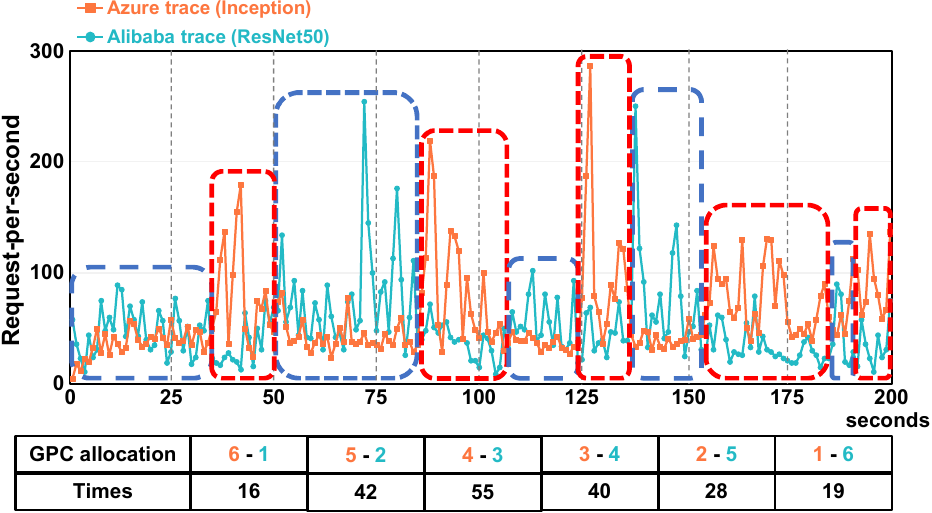}
    \caption{The figure displays the inference request per second for both the Azure and Alibaba traces within a single retraining window. The first row of the table lists GPC allocations, where `1-6' indicates Inception is allocated 1 GPCs while ResNet50 is allocated 6 GPCs. 
    }
    \label{fig:motivation_inference_varying_workload}
\end{figure}

\textbf{Fine-grain reconfiguration needed for multi-tenancy CL inferences.}
We use two real-world traces (Microsoft Azure~\cite{trace-microsoft} and Alibaba Cloud~\cite{trace-alibaba}) for Inception and ResNet50, respectively, to study the multi-tenancy CL inferences. Figure~\ref{fig:motivation_inference_varying_workload} plots the inference request arrival pattern (i.e., request per second) within a retraining window (i.e., 200 seconds). First, the arrival rate varies significantly over time within a model and between two models. For instance, during certain periods (blue boxes), ResNet50 has more inference requests than Inception, whereas for other periods (red boxes), Inception has more inference requests. Consequently, the time period in blue boxes would prefer allocations with more GPCs for ResNet50 (e.g., `3-4') to achieve higher SLO attainment, and the red box would prefer allocations with more GPCs for Inception (e.g., `4-3'). Second, the preferred allocation can vary every second as it is determined by the inference request arrival pattern. 
That is, ideally, in every second, one would need to determine the allocation that can provide the best SLO attainment based on both the current inference requests and the newly arrived inference requests.
We characterize, at each second, the GPC allocation that yields the best SLO attainment and report the number of times these GPC allocations appear during the entire retraining window in Figure~\ref{fig:motivation_inference_varying_workload}.
As shown, the GPC allocation that yields the best SLO attainment varies among all the possible allocations (i.e., from `6-1' to `1-6).
Therefore, static MIG partitioning will not work for multi-tenant CL inferences.

\begin{figure}[]
    \centering  
    \includegraphics[width=1\columnwidth]{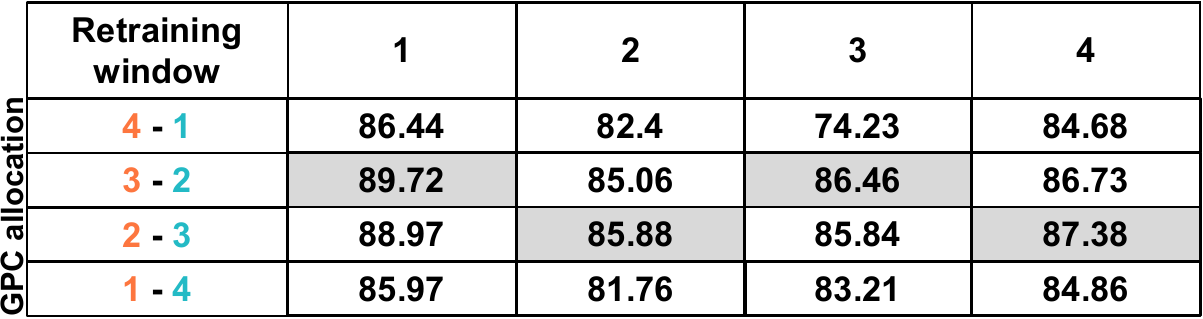}
    \caption{Inference accuracy(\%) under different GPC allocations for retraining tasks in each retraining window. The leftmost column of the table represents the GPC allocations, where `1-4' indicates that the Inception retraining task is allocated a 1-GPC instance and the ResNet50 retraining task is allocated a 4-GPC instance. The highest inference accuracy for each retraining window is highlighted.}
    \label{fig:motivation_dynamic_retraining}
\end{figure}

\textbf{Dynamic allocation needed among multi-tenancy CL retrainings.}
We further investigate the effectiveness of MIG in handling multi-tenant retraining tasks. 
To characterize the inference accuracy improvement brought by retraining, we run the inferences for both Inception and ResNet50 using the same inference trace from Microsoft Azure~\cite{trace-microsoft}, i.e., the red curve in Figure \ref{fig:motivation_inference_varying_workload}. This ensures that the improvement in inference accuracy is only affected by the retraining procedures between the two models, not the change of the inference request arrival pattern between the two models. We also fixed 1 GPC for each inference task, hence leaving 5 GPCs available for the retraining tasks between these two models. We apply a GPC allocation at the beginning of each retraining window and maintain it during that retaining window. 
Figure \ref{fig:motivation_dynamic_retraining} shows the inference accuracy results, demonstrating that no single GPC allocation consistently achieves the highest inference accuracy across all windows. 
This variability is due to the variations in model accuracy improvement across different retraining tasks. That is, the accuracy improvement through retraining tasks varies across models, and even for the same model in different retraining windows~\cite{est-acc-ekya,recommendation_1}.
As such, at each retraining window, it is beneficial to allocate more resources to the retraining task that provides a higher increase in accuracy, given that allocating more resources to a retraining task enables a shorter retraining duration and allows more inference requests to benefit from the improved model accuracy.

\section{MIGRator Design}
\label{sec:design}

We design \ourdesign, a dynamic GPU reconfiguration runtime for multi-tenancy continuous learning
workloads on modern GPUs. \ourdesign~is built upon modern MIG such that the interference among co-running tasks is eliminated. Also, \ourdesign~is aware of both inference dynamics and the retraining benefits in CL workloads. This is achieved by formulating the reconfiguration into an Integer Linear Programming (ILP) problem, which leverages a metric called {\it Goodput} (Section~\ref{sec:design_ILP_objective}) in its objective function to take into account the SLO attainment and the model accuracy. \ourdesign~uses Gurobi~\cite{gurobi} to resolve the ILP efficiently with much lower overheads compared to exhaustive search. 
\subsection{ILP Formulation}
\label{sec:design_ILP}

In this section, we present the details of formulating the MIG reconfiguration into ILP. 
The notations used in ILP formulation and their descriptions are presented in Table \ref{tab:ilp_notation}.
First, we consider the instances (denoted as $(\lambda, \gamma$)) of 12 MIG-supported configurations (denoted as $\Lambda$) illustrated in Figure \ref{fig:mig_combinations} as the basic units for GPU resource allocation. Thus, a resource allocation unit (i.e., instance) is represented by $(\lambda, \gamma)$.
These instances can be allocated to any inference (denoted as $(m, i)$) or retraining task (denoted as $(m,r)$).
The resource allocation is subject to certain constraints 
listed in Sections \ref{sec:design_ILP_constrain_general} and \ref{sec:design_ILP_constrain_task}.

The ILP solver runs at the beginning of each retraining window, to generate GPU resource allocations for each second of the entire 200-second retraining window. Thus, each retraining window has 200 possible allocations (denoted as $\Phi$).  
The ILP formulation evaluates the SLO attainment and inference accuracy (details of the objective function are provided in Section \ref{sec:design_ILP_objective}) coherently for each allocation sequence. Then, the ILP selects the allocation sequence that achieves both high SLO attainment and high inference accuracy during the entire retraining window.
Though the ILP solver is called at each retraining window, it is very efficient and completes within 1\% of the retraining window time.

\begin{table}[h]
\centering
\tiny
\begin{tabularx}{0.47\textwidth}{|>{\centering\arraybackslash}p{1.2cm}|>{\centering\arraybackslash}p{1.9cm}|X|}
\hline
 & Notation & Description \\
 \hline
\multirow{9}{=}{Meta variables} & $(m)$ & The inference task $(m,i)$ and retraining task $(m,r)$ of model $m$. \\
\cline{2-3}
& $(M,i)$ & The set of all inference tasks ($(m,i) \in (M,i)$). \\
\cline{2-3}
& $(M,r)$ & The set of all retraining tasks ($(m,r) \in (M,r)$). \\
\cline{2-3}
& $(M)$ & The set of all tasks across all models. \\ 
& & ($(M) = (M,i) \cup (M,r)$) \\
\cline{2-3}
& $\mathcal{S}$ & The retraining window size in seconds ($s \in \mathcal{S}$). \\
\cline{2-3}
& $\Lambda$ & NVIDIA MIG supported configurations~\cite{nvidia-mig-docu}. \\
\cline{2-3}
& $(\lambda, \gamma)$ & The instance $(\lambda, \gamma)$ of the configuration $\lambda$ ($\gamma \in \lambda,\ \lambda \in \Lambda)$ \\
\cline{2-3}
& $\Phi$ & A GPC allocation sequence for all tasks in current retraining window. \\ 
\hline

\multirow{16}{=}{General variables} & $X_{(m,i),(\lambda,\gamma)}^s$ & a binary variable indicating whether the instance $(\lambda,\ \gamma)$ is allocated to the inference task $(m,i)$ in the $s^{th}$ second, if the instance is allocated, the value is set to 1; otherwise, 0 \\
\cline{2-3}
 & $X_{(m,r),(\lambda,\gamma)}^s$ & a binary variable indicating whether the instance $(\lambda,\ \gamma)$ is allocated to the retraining task $(m,r)$ in the $s^{th}$ second, if the instance is allocated, the value is set to 1; otherwise, 0 \\
\cline{2-3}
& $N_{(m,i)}^s$ & The number of instances allocated to the inference task $(m,i)$ in the $s^{th}$ second. \\
\cline{2-3}
& $N_{(m,r)}^s$ & The number of instances allocated to the retraining task $(m,r)$ in the $s^{th}$ second. \\
\cline{2-3}
& $Y_{(m,i)}^s$ & The number of GPCs allocated to the inference task $(m,i)$ in the $s^{th}$ second. \\
\cline{2-3}
& $Y_{(m,r)}^s$ & The number of GPCs allocated to the retraining task $(m,r)$ in the $s^{th}$ second. \\
\hline

\multirow{3}{=}{Feasibility} & $F_{\lambda}^s$  & A binary variable indicating whether the configuration $\lambda$ is selected in $s^{th}$ second, if the configuration is selected, the value is set to 1; otherwise, 0. \\
\hline

\multirow{7}{=}{Guarantee completion and no interruption}& $C_{(m,r)}^s$  & A binary variable indicating the retraining task $(m,r)$ is running (value 1) or not (value 0) in the $s^{th}$ second. \\
\cline{2-3}
& $\mathcal{H}$ & A large constant of value 10000 to help computation in our work. \\ 
\cline{2-3}
& $RT_{(m,r)}^k$ & Retraining time of the task (m,r) in seconds given a k-GPC instance. \\
\cline{2-3}
\hline

\multirow{3}{=}{Guarantee inference task deployment} & $L_{(m, i)}$ & the minimum required size of an instance that the inference task $(m,i)$ can be deployed without error. \\
& & \\
\hline

\multirow{7}{=}{Objective related variables} & $Goouput^{\Phi}$ & The total number of valid inference requests given the GPC allocation $\Phi$ in current retraining window. \\
\cline{2-3}
& $Goouput^s_{(m,i)}$ & The total number of valid inference requests returned by task (m,i) in the $s^{th}$ second. \\
\cline{2-3}
&  $Throughput_{(m,i)}^s$ & The number of inference requests processed by task $(m,i)$ in the $s^{th}$ second. \\
\cline{2-3}
& $capability_{(m,i),(\lambda, \gamma)}$ & The number of inference requests can be processed by task $(m,i)$ given the instance $(\lambda,\gamma$. \\
\cline{2-3}
& $capability_{(m,i)}^{s}$ & The total number of inference requests can be processed by task $(m,i)$ in the $s^{th} second$. \\
\cline{2-3}
& $Recv_{(m,i)}^{s}$ & The  number of inference requests received by task $(m,i)$ in the $s^{th} second$. \\
\cline{2-3}
& {$Completion_{m}^s$} & A binary variable indicating the completion of the retraining process of model $m$ before the $s^{th}$ second, if the retraining process is completed before $s^{th}$ second, the value is set to 1; otherwise, 0. \\
\cline{2-3}
& {$accuracy_{m}^{pre}$} & The model accuracy of model $m$ before the completion of retraining process. \\
\cline{2-3}
& {$accuracy_{m}^{post}$} & The model accuracy of model $m$ after the completion of retraining process. \\
\hline

\multirow{6}{=}{Detect reconfiguration} & $R_{(m,i)}^s$ & A binary variable indicating a reconfiguration is initialized (value 1) or not (value 0) for the inference task $(m,i)$ in the $s^{th}$ second. \\
\cline{2-3}

& $\Psi_{(m,i)}$ & The average reconfiguration overhead of task $(m,i)$ in seconds during the last retraining window.\\ 
\hline

Helper variable & $\mathcal{H}$ & A large constant of value 10000 in our work. \\ 
\hline
\end{tabularx}
\caption{Symbols used in ILP.}
\label{tab:ilp_notation}
\end{table}


\subsubsection{General MIG Constraints in ILP}
\label{sec:design_ILP_constrain_general}
\hfill\\
\noindent\textbf{Ensuring valid GPU resource allocations.}
First, we must ensure that the GPU resource allocations among tasks are compatible with MIG-supported configurations (i.e., those 12 configurations in Figure \ref{fig:mig_combinations}). 
We use a binary variable $F_{\lambda}^s$ to indicate whether a MIG-supported configuration $\lambda$ is selected by any task at the $s^{th}$ second.
The $F_{\lambda}^s$ takes the value 1 (indicating being selected) only if any instance $(\lambda, \gamma)$ of this configuration is allocated to any task. This relationship can be linearized as Formula \ref{constraint:feasibility_1}, where $X_{(m), (\lambda, \gamma)}^{s}$ is a binary variable indicating whether the instance $(\lambda,\gamma)$ is allocated to model $m$'s inference or retraining task at the $s^{th}$ second.
Accordingly, the constraint to ensure the feasibility of instance allocation can be formulated as Formula \ref{constraint:feasibility_2}. 


\begin{subequations}
    \footnotesize
        \begin{align}
        &\text{Determine occupancy status for each configuration $\lambda$: }  \label{constraint:feasibility_1} \\
        &\indent F_{\lambda}^s \in \{0,1\}, \forall s, \lambda  \nonumber \\
        & \indent F_{\lambda}^s \leq \sum_{\gamma \in \lambda} \sum_{m \in M} X_{(m), (\lambda, \gamma)}^{s}, \forall s, \lambda \nonumber \\
        & \indent F_{\lambda}^s \geq \frac{1}{\mathcal{H}} \cdot \sum_{\gamma \in \lambda} \sum_{m \in M} X_{(m), (\lambda, \gamma)}^{s}, \forall s, \lambda \nonumber \\
        &\text{Ensure only one configuration $\lambda$ is selected from $\Lambda$: }  \label{constraint:feasibility_2} \\
        & \indent \sum_{\lambda \in \Lambda} F_{\lambda}^s = 1, \forall s \nonumber
        \end{align}
\end{subequations}

\noindent\textbf{Prohibit instance sharing.} 
\ourdesign~is built upon MIG and does not share MIG instances among tasks. We now formulate this into ILP constrain. 
Specifically, we use the binary variables $X_{(m,i),\ (\lambda,\ \gamma)}^{s}$ to indicate whether the instance $(\lambda,\ \gamma)$ is allocated to the task $(m,i)$ in the $s^{th}$ second.
If the instance is allocated to the task $(m,i)$, $X_{(m,i), (\lambda, \gamma)}^s$ is set to 1; otherwise, it is set to 0. The same process is also applied to $X_{(m,r), (\lambda, \gamma)}^s$.
The constraint to avoid instance sharing is formulated as Formula \ref{constraint:instance_sharing}, which ensures that an instance $(\lambda,\ \gamma)$ can be allocated to at most one task. 


\begin{equation}
    \label{constraint:instance_sharing}
    \footnotesize
        \begin{aligned}
        &\text{Prevent instance sharing: } \sum_{m \in M} (X_{(m,i), (\lambda, \gamma)}^{s} + X_{(m,r), (\lambda, \gamma)}^{s}) \leq 1, \forall s, \gamma 
        \end{aligned}
\end{equation}



\subsubsection{Task-dependent Constraints in ILP}
\label{sec:design_ILP_constrain_task}
\hfill\\
The constraints listed in Section \ref{sec:design_ILP_constrain_general} ensure that the generated GPU resource allocations are feasible and can be implemented by MIG. In this section, we list three task-dependent constraints.

\noindent\textbf{No interruption for retraining.}
\ourdesign~follows prior works~\cite{est-acc-ekya} and does not interrupt retraining tasks once they start executing, due to the high interruption overhead (including checkpointing and reloading). 
To allocate the instance for the retraining task, it follows the constraint Formula \ref{constraint:retrain_1_instance}.
As such, the necessary and sufficient condition to prevent interruptions in the retraining process is to ensure that the number of GPCs allocated ($N_{(m,r)}$) to a retraining task remains constant every second from the start of the retraining process until completion.
The constraint is formulated as follows.
Suppose a retraining task $(m,r)$ starts at the $s^{th}$ second with $Y_{(m,r)}^s$ GPCs allocated, and the time required for it to complete is $RT_{(m,r)}^{Y_{(m,r)}^s}$. 
Then we need to ensure that the number of GPCs allocated to this retraining task remains the same every second over the period $s$ to $s+RT_{(m,r)}^{Y_{(m,r)}^s}$.
This constraint is given in Formula \ref{constraint:retrain_no_interruption}.


Here, $z_{(m,r)}^s$ in Formula \ref{constraint:retrain_no_interruption} is a binary variable to indicate whether the training task $(m,r)$ is started at the $s^{th}$ second. 
$z_{(m,r)}^s$ is used to ensure that we only check the GPC number consistency within the training duration.
The definition of $z_{(m,r)}^s$ is linearly formulated in Formula \ref{constraint:retrain_start}. 
In this formulation, we also need to estimate the time needed to complete a retraining task ($RT_{(m,r)}^{Y_{(m,r)}^s}$ in Formula \ref{constraint:retrain_no_interruption}).
To achieve this, we first approximate the training latency required for a model to be three times the inference latency with the same volume of data and the same amount of GPU resources~\cite{evci2020rigging}, where the inference latency is offline profiled.
Note that a model's inference task can be allocated with instances containing different numbers of GPCs. We only need to profile the inference latency once for each specific instance, and this profiling overhead is negligible.
Then, we can calculate the training duration based on the volume of retraining data.
Besides, the function $Equals$~\cite{ilp_equals} in Formula \ref{equals_in_ilp}, is used to compare two numbers, returning 1 if they are equal and 0 otherwise.


\begin{subequations}
    \label{constraint:retrain_start_timestamp}
    \footnotesize    
    \begin{align}
        &\text{Count the number of instances allocated to task $(m,r)$: } \label{constraint:retrain_1_instance}\\ 
        & \indent N_{(m,r)}^s = \sum_{\lambda,\gamma} X_{(m,r), (\lambda,\gamma)}^s, \forall m,s \nonumber \\
        & \indent N_{(m,r)}^s \leq 1, \forall m,s \nonumber \\
        &\text{Determine the running status for task $(m,r)$: } \label{constraint:retrain_current_status} \\ 
        & \indent C_{(m,r)}^s \in \{0,1\}, \forall m,s  \nonumber\\
        & \indent C_{(m,r)}^s \leq N_{{m,r}}^s, \qquad C_{(m,r)}^s \geq \frac{1}{\mathcal{H}} \cdot N_{{m,r}}^s, \forall m, s \nonumber \\
        &\text{Determine whether task $(m,r)$ starts in $s^{th}$ second: }  \label{constraint:retrain_start} \\ 
        & \indent z_{m,r}^s \in \{0,1\}, \quad  z_{m,r}^s \leq 1 - C_{(m,r)}^{s-1}, \forall m,s \nonumber \\
        & \indent z_{m,r}^s \leq C_{(m,r)}^s, \quad z_{m,r}^s \geq C_{(m,r)}^s - C_{(m,r)}^{s-1}, \forall m,s \nonumber \\
        &\text{Count the number of GPCs allocated to task $(m,r)$: }  \\
        & \indent Y_{(m,r)}^s = \sum_{\lambda, \gamma} X_{(m,r), (\lambda, \gamma)}^{s} \cdot \text{sizeof$(\lambda, \gamma)$}, \forall m,s \nonumber \\
        &\text{Whether the same number of GPCs allocated to task $(m,r)$: }  \label{equals_in_ilp} \\
        &\indent  q_{(m,r)}^s \in \{0,1\}, \quad q_{(m,r)}^s = Equals(Y_{(m,r)}^{s-1}, Y_{(m,r)}^s), \forall m,s \nonumber  \\
        &\text{Guarantee no interruption: }   \label{constraint:retrain_no_interruption} \\ 
        &\indent z_{m,r}^s \cdot \sum_{w=s}^{s+RT_{(m,r)}^{Y_{(m,r)}^s}} q_{(m,r)}^w = RT_{(m,r)}^{Y_{(m,r)}^s}, \forall m,s \nonumber
    \end{align}
\end{subequations}

\noindent\textbf{Completing retraining tasks within the retraining window.}
As required by continuous learning application~\cite{est-acc-ekya,ctn-recl}, each retraining task must be completed within the current retraining window.
This constraint can be formulated as Formula \ref{constraint:retrain_5}.
In this formula, we first ensure that each retraining task will be launched. 
Then we guarantee the retraining task must be completed within the retraining window by ensuring its completion time is within the window.

\begin{equation}
    \footnotesize
    \label{constraint:retrain_5}
    \begin{aligned}
        &\text{Guarantee retraining task $(m,r)$ will be launched: } \\
        &\indent \sum_s C_{(m,r)}^s > 0, \forall m   \\
        & \text{Guarantee completion within the window: } \\
        &\indent z_{(m,r)}^s \cdot (s + RT_{(m,r)}^{Y_{(m,r)}^s}) \leq S, \forall m, s 
    \end{aligned}
\end{equation}

\noindent\textbf{Guarantee deployment of inference task.}
To guarantee that the model can be deployed without errors (e.g., out of memory) and serve inference requests at any time, we need to ensure that each inference task is always allocated enough resources required for deployment.
To formulate this constraint, we start by calculating the number of GPCs allocated to the inference task $(m,i)$ in the $s^{th}$ second, denoted as ($Y_{(m,i)}^s$), as shown in Formula \ref{constraint:inference_starvation_calculation_1}.
Then the constraint can be formulated as Formula \ref{constraint:inference_starvation_calculation_2}.
Specifically, for each inference task, we guarantee it is always allocated with at least one $L_{(m,i)}$-GPC instance, where the $L_{(m,i)}$-GPC instance is the minimum resource demand for this inference task to be launched without error.


\begin{subequations}
\footnotesize
\begin{align}
    &\text{Count the number of GPCs allocated to task $(m,i)$: }  \label{constraint:inference_starvation_calculation_1} \\
    & \indent Y_{(m,i)}^s = \sum_{\lambda, \gamma} X_{(m,i), (\lambda, \gamma)}^{s} \cdot \text{sizeof$(\lambda, \gamma)$}, \forall m,s  \nonumber \\
    &\text{Guarantee deployment for task $(m,i)$: }\ Y_{(m,i)}^s \geq L_{(m,i)},\ \forall m, s \label{constraint:inference_starvation_calculation_2} 
\end{align}
\end{subequations}



\subsubsection{Objective Function.}
\label{sec:design_ILP_objective}
\hfill\\
Our optimization goal is to ensure that the generated allocation sequence achieves both high SLO attainment and high inference accuracy over the entire retraining window.
Neither SLO attainment nor inference accuracy alone can fully reflect system performance.
Therefore, to assess system performance more accurately, we leverage the metric $Goodput$, as shown in Equation \ref{equation:perf_metric_model_level}.
The $Goodput$ is calculated as follows. For each inference request arrived during a given retraining window, it is considered a ``valid'' inference request only if the request satisfies two conditions: i) timeliness (SatisfySLO) and ii) correctness (SatisfyPrediction). 
Specifically, timeliness indicates the completion time of inference request meets the SLO target (i.e., the output of function SatisfySLO is ``1'' if timeliness is met, otherwise ``0'').
Correctness represents the request's inference outcome is correct (i.e., the output of function SatisfyPrediction is ``1'' if the returned result is correct, otherwise ``0'').
Both conditions are binaries for a request. Then, the $Goodput$ counts the total number of ``valid'' inference requests that satisfy the above two conditions. Therefore, a higher $Goodput$ indicates that more inference requests that contributes to SLO attainment and also leverage the updated model. The equation of $Goodput$ is given as follows: 

\begin{equation}
 \footnotesize
\label{equation:perf_metric_model_level}
    \begin{aligned}
    \text{Goodput} &= \sum\limits_{r} \text{SatisfySLO}(r) \wedge \text{SatisfyPrediction}(r), \forall  r \in \substack{\text{inference}\\ \text{requests}}
    \end{aligned}
\end{equation}

Recall our discussion at the beginning of Section~\ref{sec:design_ILP}, at the start of each retraining window, \ourdesign~generates GPC allocations on a per-second basis for the entire given retraining window. Subsequently, \ourdesign~leverages ILP to evaluate all the GPC allocations collectively as a GPC allocation sequence within a retraining window. We denote this GPC allocation sequence as \(\Phi\).
As such, the objective function of our ILP is to evaluate the generated allocation sequences and identify the one that maximizes the number of valid inference requests satisfying both timeliness and correctness requirements:
\begin{equation}
    \footnotesize
    \label{equation:ilp_objective}
    \textbf{Maximize} [Goodput^{\Phi}] 
\end{equation}
Here, $Goodput^{\Phi}$ is calculated by summing the number of valid inference requests from each model's inference task (denoted as $(m,i)$) in each second (denoted as $s$) within the retraining window, which can be expressed as:
\begin{equation}
    \footnotesize
    \label{equation:ilp_objective2}
    Goodput^{\Phi} = \sum_{m, s} Goodput_{(m,i)}^{s}
\end{equation}
Specifically, $Goodput_{m,i}^s$
is calculated by multiplying the number of inference requests processed for task $(m,i)$ in the $s^{th}$ second (denoted as $Throughput_{(m,i)}^s$) by the model accuracy: 
\begin{equation}
    \footnotesize
    \begin{aligned}
    Goodput_{(m,i)}^s = & Throughput_{(m,i)}^s \cdot [(1-Completion_{m}^s) \cdot accuracy_{m}^{pre} \\
    &\indent\indent\indent\indent\indent\indent + Completion_{m}^s \cdot accuracy_{m}^{post}] 
    \end{aligned}
\end{equation}
In this equation, $accuracy_{m}^{pre}$ and $accuracy_{m}^{post}$ indicates the pre-retraining and post-retraining accuracy of model $m$, respectively.
Given the throughput of the inference task at the second, there can be two situations.
First, if the model retraining process is not finished yet at this second $s$, giving $Completion_{m}^s=0$, then inference requests are served with the model before retraining, denoted as $accuracy_{m}^{pre}$ in the objective function.
Second, if the model retraining process is finished before this second $s$, giving $Completion_{m}^s=1$, then the updated model is available to serve inference requests, denoted as $accuracy_{m}^{post}$ in the objective function.

In the next section \ref{sec:design_ILP_estimation}, we will introduce how we calculate the $Throughput_{(m,i)}^s$, retraining task status $Completion_{m}^s$, and the post-retraining accuracy $accuracy_{m}^{post}$ in the objective function. 

\subsubsection{Calculating Parameters in Objective Function.}
\label{sec:design_ILP_estimation}
\hfill\\
\textbf{Calculating} \bm{$Throughput_{(m,i)}^s$}.
To estimate how many inference requests of task $(m,i)$ are processed in the $s^{th}$ second (i.e., $Throughput_{(m,i)}^s$), we need to know i) the total number of inference requests of task $(m,i)$ in the \( s^{th} \) second (denoted as $Recv_{(m,i)}^s$), and ii) the maximum number of inference requests that can be processed for task $(m,i)$ within the \( s^{th} \) second (denoted as $capability_{(m,i)}^s$). 
Then the $Throughput_{(m,i)}^s$ is the minimum of $Recv_{(m,i)}^s$ and $capability_{(m,i)}^s$, which is expressed as: $Throughput_{(m,i)}^s = minimum\{Recv_{(m,i)}^{s}, capability_{(m,i)}^s\}$. Here we leverage the formula $minimum(A,B)$ in~\cite{ilp_equals} which takes two numbers $A$ and $B$ and returns the minimum value between them.

\begin{equation}
\label{objective:throughput}
\footnotesize
\begin{aligned}
            &\text{capability of $(m,i)$ in $s^{th}$ second}: \\
            & \indent capability_{(m,i)}^s = \sum_{\lambda, \gamma} X_{(m,i), (\lambda, \gamma)}^{s} \cdot capability_{(m,i),(\lambda, \gamma)} \\
            & \indent\indent\indent\indent\indent - \frac{R_{(m,i)}^s \cdot \Psi_{(m,i)}}{\sum_{\lambda, \gamma} X_{(m,i), (\lambda, \gamma)}^{s}} \cdot  capability_{(m,i),(\lambda, \gamma)} \\
        \end{aligned}
\end{equation}

To obtain the $Recv_{(m,i)}^s$, at the beginning of the retraining window, we predict the number of inference requests arriving every second throughout the entire window based on the historical inference request arrival data from previous windows.
Specifically, we follow previous works~\cite{informer-gong2022load,informer-hua2023kae,informer-wu2022long} and leverage a transformer-based model Informer~\cite{informer-itself} for prediction. Informer is widely used due to its effectiveness and accuracy in long-sequence forecasting, and its efficiency for real-time prediction.
Regarding $capability_{(m,i)}^s$, we sum up the number of inference requests that can be processed by each instance allocated to task $(m,i)$ (denoted as $capability_{(m,i),(\lambda,\gamma)}$), where $(\lambda, \gamma)$ represents an instance, as shown in Equation \ref{objective:throughput}.

\begin{figure}
    \centering  
    \includegraphics[width=.48\textwidth]{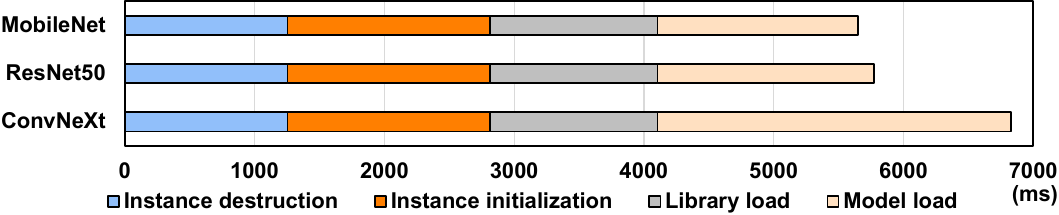}
    \caption{Reconfiguration overhead for ConvNeXt, ResNet50 and MobileNet.}
    \label{fig:design_reconfig_cost}
\end{figure}

\textbf{Taking reconfiguration overhead into account.} So far, we have not included the reconfiguration overhead in our ILP formulation. However, in reality, there are considerable overheads associated with MIG reconfiguration. This reconfiguration overhead includes both the application aspect (e.g., model initialization and parameter loading) and the hardware aspect (e.g., driver handling the reconfiguration).
Figure~\ref{fig:design_reconfig_cost} plots our characterization of three models and their reconfiguration overheads. As one can observe, the reconfiguration overheads can take more than 6,000 milliseconds, which is significantly longer — more than 1,000 times — than the latency of responding to a single inference request. Therefore, ignoring this reconfiguration overhead during scheduling could lead to frequent reconfigurations, resulting in more operational overheads than the benefits derived from resource reconfiguration.
Since the instances that are being reconfigured cannot serve any task, the reconfiguration overheads inevitably affect the inference throughput.
We take this into account by Equation \ref{objective:throughput}. $\Psi_{t_i}$ in the equation represents the time needed for reconfiguration, such as instance initialization and model loading.
Specifically, when there is a MIG reconfiguration in the $s^{th}$ second and task $(m,i)$ is affected by this reconfiguration, the throughput of the task $(m,i)$ will inevitably drop due to the reconfiguration time.

\textbf{Detecting Reconfiguration.} We use a binary variable $R_{(m,i)}^s$ to indicate whether a task $(m,i)$ is affected by MIG reconfiguration.
The $R_{(m,i)}^s$ takes the value 1 (indicating task $(m,i)$ being affected) either when the number of instances allocated ($N_{(m,i)}^s$) or the number of GPCs allocated ($Y_{(m,i)}^s$) to $(m,i)$ changes. 
The $R_{(m,i)}^s$ is linearly formulated as:
\begin{equation}
    \label{constraint:reconfiguration}
    \footnotesize
    \begin{aligned}
            &\text{Reconfiguration variable}: \\
            & \indent R_{(m,i)}^s \in \{0,1\} \\
            &\text{Is the same number of GPCs allocated: } \\
            & \indent equalGPC_{(m,i)}^s = Equals(Y_{(m,i)}^{s-1}, Y_{(m,i)}^s) \\
            &\text{Is the same number of instances allocated: } \\
            & \indent equalInst_{(m,i)}^s = Equals(N_{(m,i)}^{s-1}, N_{(m,i)}^s) \\
            &\text{Determine if a reconfiguration is initiated}: \\
            & \indent   R_{(m,i)}^s \leq equalGPC_{(m,i)}^s + equalInst_{(m,i)}^s, \forall m, s \\ 
    \end{aligned}
\end{equation}
Here, we leverage $Equals$~\cite{ilp_equals} to determine whether two numbers are equal, the same process we did in Formula~\ref{equals_in_ilp}. If two numbers are equal, $Equals$ returns value 1; otherwise, 0.

\textbf{Detecting the completion of a retraining process \bm{$Completion_m^s$} and estimating post-retraining model accuracy.}
Given our constraint (Formula \ref{constraint:retrain_start_timestamp} and \ref{constraint:retrain_5}) that ensures once a retraining task $(m,r)$ is launched, it must not be interrupted until completion, our method to determine if a retraining task is complete is based on the running status of $(m,r)$. Specifically, we consider a retraining task to be completed during the $(s-1)^{th}$ second if it is running at that second but not at the subsequent $s^{th}$ second.
The running status of retraining task $(m,r)$ in the $s^{th}$ second is represented by a binary variable $C_{(m,r)}^{s}$, where $C_{(m,r)}^{s} = 1$ indicates that $(m,r)$ is actively running. The detailed definition of $C_{(m,r)}^{s}$ is provided in Formula \ref{constraint:retrain_current_status}.
To explicitly indicate the completion status of $(m,r)$, we employ a binary variable $Completion_{m}^s$. When $Completion_{m}^s = 1$, it indicates that the retraining task $(m,r)$ has been completed before the $s^{th}$ second. The determination process for $Completion_{m}^s$ is linearly formulated in Formula \ref{constraint:retrain_end}.
To quickly estimate the post-retraining model accuracy, we adopt the methodology used in prior works~\cite{est-acc-optimus,est-acc-ekya}.
Specifically, we sample a small subset of the retraining data and use it to train the model for a few epochs. This preliminary retraining allows us to obtain a model accuracy improvement curve. Based on this curve, we are able to predict the model's accuracy upon convergence when trained with the complete dataset.

\begin{footnotesize}
    \begin{equation}
    \label{constraint:retrain_end}
        \begin{aligned}
            &\text{Retraining process finished in previous second: } \\
            &\indent k_{(m,r)}^s \in \{0,1\}, \quad  k_{(m,r)}^s \geq C_{(m,r)}^{s-1} - C_{(m,r)}^{s}, \forall m,s\\
            & \indent k_{(m,r)}^s \leq C_{(m,r)}^{s-1}, \indent k_{(m,r)}^s \leq 1 - C_{(m,r)}^s, \forall m,s \\
            &\text{Retraining completion status: } \\
            &\indent Completion_{(m,r)}^s \in \{0,1\}, \forall m, s \\ 
            &\indent Completion_{m}^s  \geq Completion_{m}^{s-1}, \forall m, s \\
            & \indent Completion_{m}^s \geq  k_{(m,r)}^s, \forall m, s \\
            & \indent Completion_{m}^s \leq  Completion_{m}^{s-1} + k_{(m,r)}^s, \forall m,s \\
        \end{aligned}
    \end{equation}
\end{footnotesize}

After calculating $Throughput(m,i)^s$, the retraining task status $Completion_m^s$ and the post-retraining model accuracy $accuracy_m^{post}$, we can evaluate $Goodput$ for GPC allocations and identify the beneficial GPC allocation $\Phi$ for the entire retraining window.
To address these complex resource allocation challenges efficiently, \ourdesign~employs the Gurobi solver\cite{gurobi}, a powerful optimization tool known for its efficiency and effectiveness in solving large-scale linear programming problems.

\subsection{Efficient Transition to Optimal GPC Allocation}
\label{sec:design_eaa}




So far, we have not explored any optimizations to reduce the reconfiguration overhead. In practice, if one can reduce the reconfiguration overheads, it can enable the ILP solver to find more beneficial GPC allocations. To this end, we propose a novel technique `pre-initialization' that effectively reduces the reconfiguration overheads ($\Psi_{(m,i)}$ in Equation \ref{objective:throughput}). Note that, reducing the overheads will not affect the ILP constraints.

\begin{figure}[tbp]
    \centering  
    \includegraphics[width=.49\textwidth]{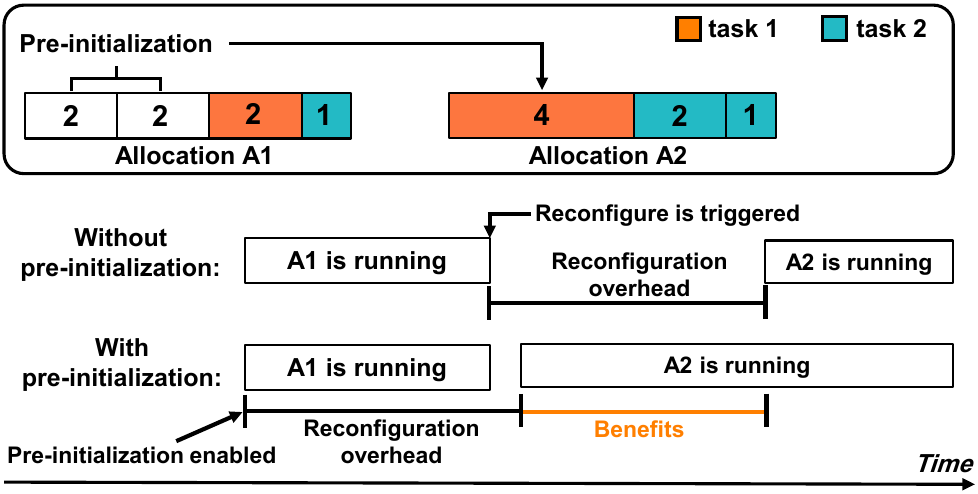}
    \caption{An example of pre-initialization to reduce reconfiguration overheads.}
    \label{fig:design_eaa_workflow}
\end{figure}


We illustrate the overhead reduction strategy using an example depicted in Figure \ref{fig:design_eaa_workflow}. 
We assume there are two co-located tasks in the example. A1 and A2 represent two consecutive GPU resource allocations generated by ILP for these tasks. Specifically, A1 requires allocating one 2-GPC instance to task 1 and one 1-GPC instance to task 2. A2 changes the allocation to one 4-GPC instance for task 1 and one 2-GPC instance along with one 1-GPC instance for task 2. 



Comparing allocations A1 and A2, one can observe that two unused 2-GPC instances in A1 can be combined to constitute the 4-GPC instance in A2 before reaching A2. Pre-initializing this 4-GPC instance can overlap the reconfiguration process with the computation before reaching A2, thus hiding the reconfiguration overhead between A1 and A2.
This allows the tasks to use the reconfigured instances sooner, as shown in Figure \ref{fig:design_eaa_workflow}.
Note that we only leverage the unused instance for `pre-initialization', so that it will not affect any tasks running on occupied instances.
After using ILP to obtain the whole sequence of GPU resource allocations for the entire retraining window, \ourdesign~will quickly traverse the allocations sequence to identify opportunities for pre-initialization.

\section{Evaluation}
\label{experiment}

\subsection{Experimental Setup}
\label{experiment_setup}
\noindent\textbf{System:}
We perform experiments on NVIDIA A100 GPU with 40 GB memory and CUDA 12.1. The GPU driver version is 550.54.15, which supports both MPS and MIG GPU allocation.  The server installs AlmaLinux-8.7 OS. 


\noindent\textbf{DNN models:}
We use six models with different GFLOPs listed in Table \ref{tab:model_table}, and each model is labeled as either `High' or `Low' based on its GFLOPs. 

\noindent\textbf{Inference request traces:}
\label{sec:experiment_inference_workload}
We use two real-world inference traces for inference tasks. One is derived from Alibaba Cluster trace~\cite{trace-alibaba}, whereas the other is from the Microsoft Azure trace~\cite{trace-microsoft}. The trace characteristics are given early in Figure~\ref{fig:motivation_inference_varying_workload}. For the main results, we report the typical inference batch size (i.e., 1). We also report a larger inference batch size (i.e., 4) in Section~\ref{sec:sensitivity}. 


\noindent\textbf{Retraining datasets:}
\label{retraining_benchmarks}
We utilize three widely-used continuous learning retraining datasets: NC-CIFAR-10~\cite{cifar10}, NC-CORe50~\cite{core50}, NC-20-Newsgroups~\cite{twentyNews}. Specifically, in NC-CIFAR10, there are a total of 5 scenarios of training data, and each scenario introduces two new data classes. The first scenario training data is used to pre-train the model, and the rest four scenario training data is used to retrain the model (corresponding to four retraining windows). 
In NC-CORe50, there are 50 classes. Models are pre-trained using the first five classes. Then, in each retraining window, five additional classes are introduced for retraining, leading to nine retraining windows.
In NC-20-Newsgroups, there are 20 classes. Models are pre-trained using the first two classes. Two new classes are added in each retraining window, resulting in a total of nine retraining windows.
In our experiments, NC-20-Newsgroups is used for the language model Bert, while NC-CIFAR-10 and NC-CORe50 are used for vision models (those 5 DNN models in Table~\ref{tab:model_table} except for Bert). 

\noindent\textbf{Workloads:}
The evaluation includes 16 workloads listed in Table \ref{tab:model_combinations}, where each workload consists of two co-running tenants (i.e., two CL models). The workloads are differentiated by different models, different retraining data sets, and different inference traces. We cover diverse combinations to show the generalizability of our design. 

\begin{table}[t]
    \renewcommand\arraystretch{1.2}
    \footnotesize
    \centering
    \begin{tabular}{|c|c|c|} \hline
        Model & GFLOPs & Abbreviation\\ 
        \hline
         Bert$_{base}$ & 22.2G (High)  & Bert \\ 
        \hline
         ViT$_{b\_16}$ & 17.56G (High) &ViT \\ 
        \hline
         ConvNeXt$_{base}$ & 15.36G (High) & ConvNeXt \\ 
        \hline
         Inception$_{v3}$ & 5.71G (Low) & Inception \\ 
        \hline
         ResNet$_{50}$ & 4.09G (Low) & ResNet50 \\ 
        \hline
         MobileNet$_{v2}$ & 0.32G (Low) & MobilNet\\ 
        \hline
    \end{tabular}
    \caption{List of models.}
    \vspace{-2pt}
    \label{tab:model_table}
\end{table}

\begin{table}[t]
\centering
\footnotesize
\setlength{\tabcolsep}{0.3pt}
\begin{tabular}{|l|l|l|l|l|l|l|}
\hline
{\bf Work} & {\bf Model-1} & {\bf Inference} & {\bf Retraining} &{\bf Model-2} & {\bf Inferece} & {\bf Retraining}\\
{\bf -load} &  & {\bf trace}& {\bf dataset} & &{\bf trace}  & {\bf dataset}\\

\hline
W1 & Bert & Alibaba & NC-20N & ViT & Azure & NC-CIF  \\
\hline
W2 & Bert & Alibaba & NC-20N &  ConvNeXt & Azure &NC-CIF  \\
\hline
W3 & ViT & Alibaba & NC-CIF &  ConvNeXt & Azure & NC-CIF  \\
\hline
W4 & Bert & Alibaba & NC-20N &  Inception & Azure &NC-CIF   \\
\hline
W5 & ViT & Alibaba & NC-CIF &  ResNet50 & Azure & NC-CIF  \\
\hline
W6 & ConvNeXt & Alibaba & NC-CIF &  MobileNet & Azure & NC-CIF  \\
\hline
W7 & Inception & Alibaba & NC-CIF &  ResNet50 & Azure &NC-CIF   \\
\hline
W8 & ResNet50 & Alibaba & NC-CIF &  MobileNet & Azure &NC-CIF   \\
\hline
W9 & Bert & Alibaba & NC-20N & ViT & Azure & NC-COR \\
\hline
W10 & Bert & Alibaba & NC-20N &  ConvNeXt & Azure & NC-COR\\
\hline
W11 & ViT & Alibaba & NC-COR &  ConvNeXt & Azure & NC-COR  \\
\hline
W12 & Bert & Alibaba & NC-20N &  Inception & Azure & NC-COR  \\
\hline
W13 & ViT & Alibaba & NC-COR &  ResNet50 & Azure & NC-COR  \\
\hline
W14 & ConvNeXt & Alibaba & NC-COR &  MobileNet & Azure & NC-COR  \\
\hline
W15 & Inception & Alibaba & NC-COR &  ResNet50 & Azure & NC-COR  \\
\hline
W16 & ResNet50 & Alibaba & NC-COR &  MobileNet & Azure & NC-COR  \\
\hline
\end{tabular}
\caption{Experitmental multi-tenancy workloads. In the retraining datasets, ``NC-CF10'', ``NC-COR'', and ``NC-20N'' represents the NC-CIFAR-10~\cite{cifar10}, NC-CORe50~\cite{core50}, NC-20-Newsgroups~\cite{twentyNews} datasets, respectively.}
\vspace{-2pt}
\label{tab:model_combinations}
\end{table}




The retraining window size is set to 200 seconds and fixed through all experiments, aligning with prior continuous learning works\cite{est-acc-ekya, ctn-recl}.
We compare MIGRator with multiple state-of-the-art GPU resource allocation works:
\squishlist{}
    \item \textbf{Astraea~\cite{spatialmps-astraea}} Astrea is a MPS-based GPU resource management framework. It dynamically allocates SMs among tasks to improve QoS and resource efficiency.
    \item \textbf{PARIS~\cite{spatialmig-paris}} PARIS and ELSA is a MIG-based work, which statically partitions GPCs based on the model's computing intensity.
    \item \textbf{Ekya~\cite{est-acc-ekya}} Ekya is a MPS-based continuous learning the-state-of-art work. Ekya adjusts SM allocations for both inference and retraining tasks at the start and end of retraining processes.
\squishend{}


\subsection{Experimental Results}
\label{sec:evaluation_explanation}

\begin{figure}[t]
    \centering  
    \includegraphics[width=\columnwidth]{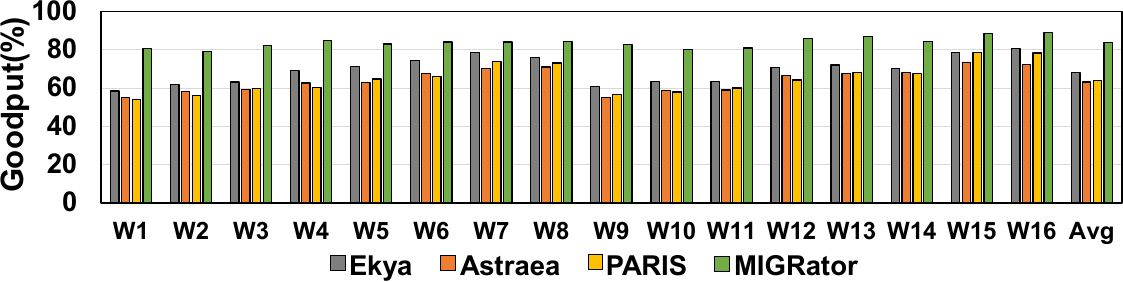}
    \caption{$Goodput$(\%) of different workloads.}
    \label{fig:evaluation_batch_1}
\end{figure}

\textbf{Goodput evaluation.} For each workload, we compute the $Goodput$ (defined in Formula \ref{equation:perf_metric_model_level} in Section~\ref{sec:design_ILP_objective}) of each retraining window, and sum all the $Goodput$ across all the retraining windows together as the overall execution $Goodput$. We normalize the overall execution $Goodput$ to the total number of inference requests during the entire execution and report the percentage results in Figure~\ref{fig:evaluation_batch_1}. The higher the percentage, the more number of valid inference requests (i.e., better performance). From the results, one can observe that \ourdesign~significantly and uniformly improves the overall $Goodput$ compared to existing state-of-the-art approaches.  Specifically, \ourdesign~improves the $Goodput$ by an average of 17\%, 21\%, and 20\% compared to Ekya, Astraea, and PARIS, respectively. 
This is due to the capability that \ourdesign~determines the GPC allocation to coherently improve both the SLO attainment and the inference accuracy at a per-second granularity. The detailed reason behind the improvements comes from multiple aspects, including i) inference accuracy improvement, ii) SLO attainment improvement, and iii) co-running tenant interference reduction.  We next elaborate on each in detail.

\begin{figure}[t]
    \centering  
    \includegraphics[width=\columnwidth]{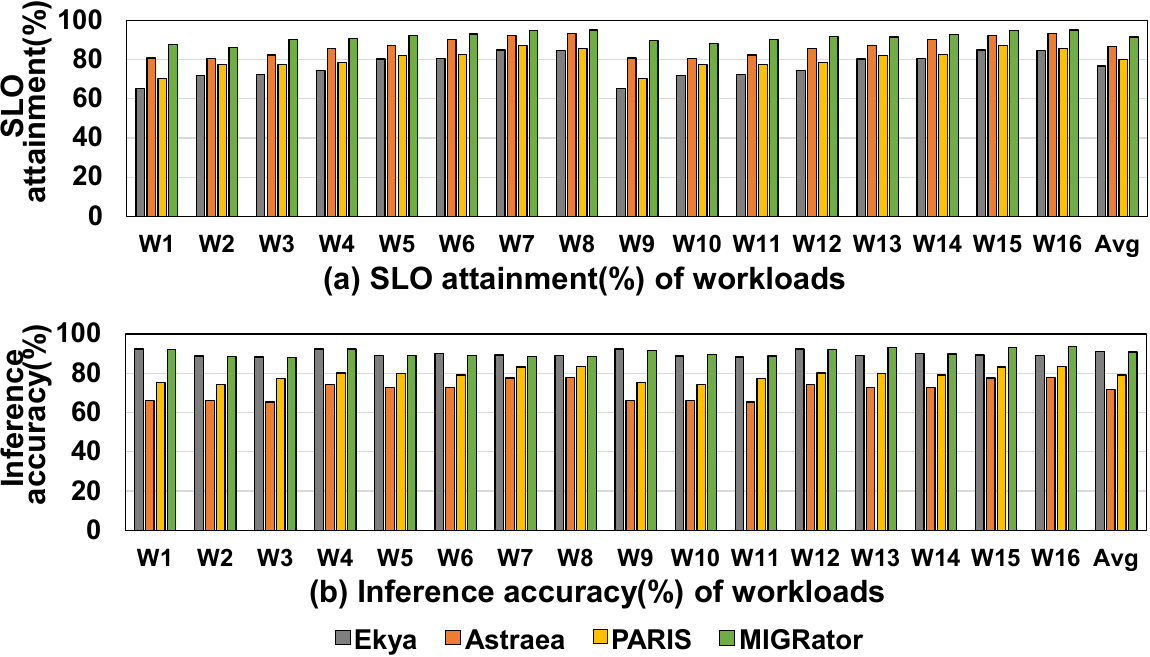}
    \caption{SLO attainment(\%) and inference accuracy(\%).}
    \label{fig:evaluation_single_supplementary}
\end{figure}

\textbf{Inference accuracy improvements. }We show the inference accuracy improvements in Figure \ref{fig:evaluation_single_supplementary}(b). As one can observe, \ourdesign~improves the average inference accuracy by 19\%, and 12\%, over Astraea, and PARIS. Recall that Astraea is an MPS-based approach that conducts the resource allocation based on the compute intensity. While it could allocate more resources to retraining tasks based on the retraining compute intensity, it is not aware of the benefits (i.e., inference accuracy improvements) brought by the retraining. For PARIS, it is a MIG-based approach that statically partitions the GPUs based on computing intensity; it neither allows reconfiguration during execution nor is it aware of the benefits brought by retraining. In contrast, our approach performs dynamic MIG reconfiguration and, most importantly, values the significance of accuracy improvements when doing retraining in each restraining window. That is, if the retraining is able to improve the model accuracy significantly, it is likely to have more resource allocation in our approach (e.g., considered in the ILP objective function~\ref{equation:ilp_objective}). On the other hand, if the accuracy improvement is not significant, our approach could provide more resources for SLO attainment. For instance, under workload W1, \ourdesign~improves the average inference accuracy by 27\% and 17\% compared to Astraea and PARIS.
This is because models (Bert and ViT) encounter a significant model accuracy drop (by an average of 32.5\%) when new classes are introduced, whereas retraining tasks boost accuracy by approximately 30\%. In such scenarios, allocating more resources to accelerate the retraining tasks can greatly improve inference accuracy. Meanwhile, \ourdesign~achieves similar inference accuracy compared to Ekya. This is because Ekay also considers the retraining benefits (i.e., accuracy improvements) in resource allocation. It always guarantees and prioritizes enough resources for retraining when the benefits are significant. However, Ekay is not able to handle inference SLO attainment based on its current ``always-retraining'' approach, which we will elaborate on next. 

\textbf{SLO attainment improvements. }Figure~\ref{fig:evaluation_single_supplementary}(a) shows the SLO attainment achieved by \ourdesign~and other comparison approaches. On average, \ourdesign~outperforms Ekya, Astraea, and PARIS, by 17\%, 6\%, and 13\%, respectively. Specifically, Ekya and PARIS only conduct resource reconfiguration at i) the beginning of the retaining process and ii) the end of the retraining process within a given retraining window. Therefore, Ekya and PARIS cannot effectively respond to the dynamic inference request arrival pattern (generally at fine granularity on a second basis), thus negatively impacting SLO attainment. Moreover, Ekya uses exhaustive searches to find the beneficial configuration. The overheads with exhaustive searches prevent it from exploring the configuration on a per-second basis. As a result, it cannot accommodate the inference arrival pattern at fine granularity.  PARIS doesn't consider the dynamic GPC reconfiguration over time.
In contrast, \ourdesign~proactively considers the inference request arrival pattern at fine granularity and leverages ILP to determine potential GPC allocations on a per-second basis, thereby delivering higher SLO attainment. \ourdesign~also outperforms Astraea by 6\%. While both  \ourdesign~and Astraea enable dynamic resource allocations for inference tasks, Astraea is an MPS-based approach that only partitions the computing resources (i.e., SMs), leaving the memory bandwidth under contention. This is particularly problematic for large models. In contrast, \ourdesign~leverages MIG to partition both compute and memory resources among co-located tasks, ensuring no interference and maintaining high SLO attainment. 
For instance, in workload W3, \ourdesign~improves SLO attainment by 8\%, with an average improvement of 6\% across all model combinations, compared to Astraea.

\textbf{Reconfiguration overhead reduction.} Recall our discussion in Section~\ref{sec:design_eaa}, we propose pre-initialization to reduce the reconfiguration overhead. We evaluated this optimization and observed that it achieves an 83\% reduction in MIG reconfiguration overheads.

\begin{figure}[]
    \centering  
    \includegraphics[width=\columnwidth]{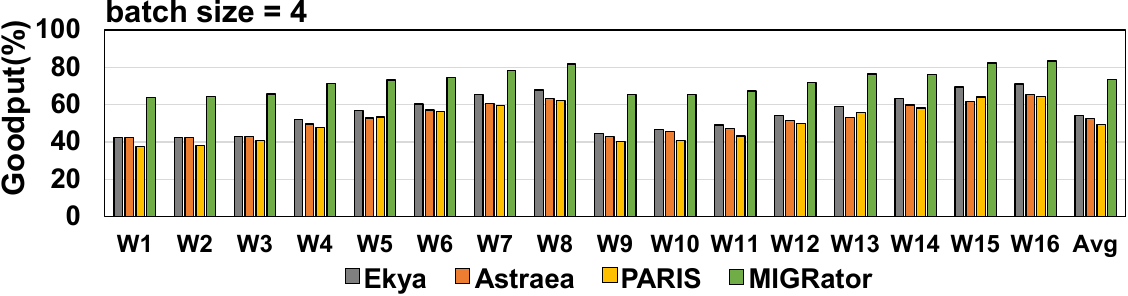}
    \caption{$Goodput$(\%) of workloads when inference batch size is 4.}
    \label{fig:evaluation_goodput_bs4}
\end{figure}



\begin{figure}[t]
    \centering  
    \includegraphics[width=1\columnwidth]{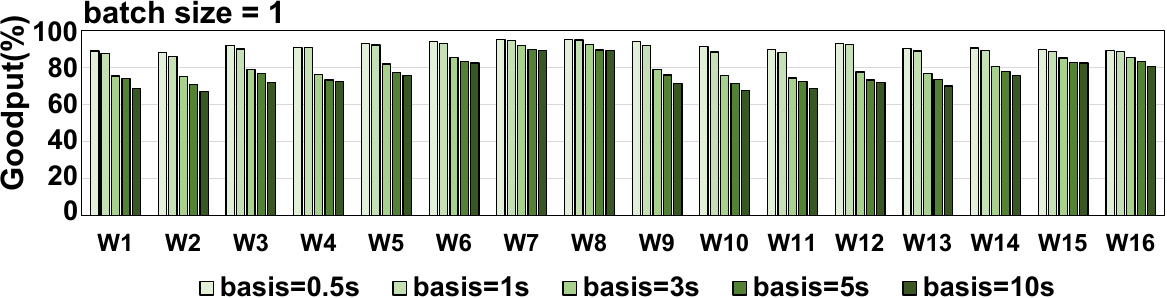}
    \caption{$Goodput$(\%) of workloads under the basis of various seconds.}
    \label{fig:single_sweep_nc_cifar10}
\end{figure}



\subsection{Sensitivity}
\label{sec:sensitivity}

\textbf{Large inference batch size. }Figure~\ref{fig:evaluation_goodput_bs4} shows the $Goodput$ of all workloads when the inference batch size is 4 for all the inference tasks. As one can observe, \ourdesign~outperforms prior works in all workloads.  The reason behind the improvements is due to the improved SLO attainment and inference accuracy, similar to the scenario when batch size is 1 (Section~\ref{sec:evaluation_explanation}). This indicates that \ourdesign~can automatically determine the beneficial GPC allocations for different batch sizes and provide additional benefits.

\textbf{Reconfiguration granularity. }So far, our discussion has focused on the reconfiguration on a per-second basis. We next report the results of \ourdesign~reconfiguration using different time granularity. Figure~\ref{fig:single_sweep_nc_cifar10} plots the results when we vary the granularity from 0.5 seconds to 10 seconds. 
As one can observe,  0.5-second granularity achieves the highest $Goodput$ across all workloads. This is because a finer granularity allows for more frequent GPC reconfigurations, better capturing the inference dynamics in continuous learning. In contrast, the 10-second granularity has a significantly lower $Goodput$ due to its limited reconfiguration frequency. However, it is worth mentioning that a finer granularity leads to more search space and higher ILP overhead. We found that 1-second granularity provides comparable $Goodput$ with 0.5-second granularity while reducing the ILP solver overheads. 
\section{Related works}
\textbf{Continuous learning.} 
Continuous learning research utilizes various strategies to maintain model accuracy over time. Some approaches focus on facilitating accurate and timely scenario change detection, which involves tracking the distribution of incoming data and detecting data drifts~\cite{wu2022energy, liu2020energy, liu2020energy}. 
These approaches enable timely model updates once the model deployment scenario changes (indicated by data drift).
Some other approaches aim to find better model retraining hyperparameters (e.g., adaptive learning rates) to accelerate model convergence~\cite{finetune_1,finetune_3} and to improve post-retraining model accuracy~\cite{finetune_2,finetune_4}. 
Moreover, some methods propose to filter out unimportant incoming retraining data to reduce the training time without compromising accuracy~\cite{panda2016conditional, wu2021enabling}, which could enables more inference requests benefits from the updated model.
It is important to emphasize that these approaches are complementary to our design. We focus on dynamic and effective resource allocation among multi-tenancy inference and retraining tasks in continuous learning, addressing the dynamic nature of both retraining and inference resource demands.


\textbf{GPU multi-tenant system}. 
Optimizing GPU resource allocation and utilization in large-scale computing environments is a popular research area. Gpulet~\cite{spatialmps-servingheterogeneousATC} leverages NVIDIA MPS to partition resources among co-running tasks, enhancing utilization and throughput. However, Gpulet adopts a coarse-grained resource reallocation strategy (i.e., reallocate resources every 20 seconds) due to the high overhead of frequent recourse reallocation. Such a coarse granularity is insufficient to effectively handle the dynamics in continuous learning applications.
Another GPU resource allocation work INFless~\cite{spatialmps-infless}, which uses NVIDIA MPS to facilitate device sharing, adopts a finer-grained resource scheduling strategy. However, it lacks optimizations to mitigate MPS reconfiguration overheads, which could significantly degrade the system performance when frequent MPS reconfiguration is needed.
In summary, while Gpu-let and INFless offer valuable approaches to GPU resource management, they have limitations. Our \ourdesign~addresses these by ensuring an efficient, fine-grained, and low-overhead resource reallocation to meet the dynamics of continuous learning applications.

\section{Conclusion}
In this paper, we propose \ourdesign, a dynamic GPU reconfiguration runtime for multi-tenancy continuous learning workloads on modern GPUs. \ourdesign~leverages MIG to minimize the interference among co-running tasks and enables fine-granular reconfigurations to improve GPU resource utilization. We formulate the reconfiguration optimization into an Integer Linear Programming (ILP) problem to take into account both the SLO attainment and retraining benefits in CL workloads. Experimental results indicate our approach significantly outperforms the state-of-the-art GPU sharing approaches for multi-tenant continuous learning workloads. 



\end{document}